# Quantum Image Preparation Based on Exclusive Sum-of-Product Minimization and Ternary Trees

M. Younatan, S. Ghose
Dept. of Physics and Computer Science
Wilfrid Laurier University
Waterloo, ON, Canada
matyounatan@gmail.com

*Abstract* — **Quantum image processing is one of the promising fields of quantum information. The complexity overhead to design circuits to represent quantum images is a significant problem. So, we proposed a new method to minimize the total number required of quantum gates to represent the quantum image. Our approach uses ternary trees to reduce the number of Toffoli gates in a quantum image circuit. Also, it uses the complement property of Boolean algebra on a set of Toffoli gates to combine two Toffoli gates into one, therefore reducing the number of overall gates. Ternary trees are used to represent Toffoli gates as they significantly increase run time and is supported through experiments on sample images. The experimental results show that there is a high-speed up compared with previous methods, bringing the processing time for thousands of Toffoli gates from minutes to seconds.**

## 1 Introduction

Quantum computers are making headlines across the world, starting with Feynman, who first proposed the idea of quantum computers in 1982 [1]. The first theoretical framework for a quantum computer was proposed in [2]. Afterward, various experimental demonstrations of quantum superposition [3] and many other useful properties of quantum mechanics have been developed [4]. Currently, these quantum computers are limited to the number of qubits and mainly act as a benchmark for small-scale algorithms [5]. These algorithms, such as image representation, use too many qubits and primitive gates to run on these machines and are prone to control-errors and decoherence [5]. Having a fault-tolerant universal quantum computer is years or decades away [6]. The performance of quantum algorithms measured by the number of primitive gates employed to construct a quantum circuit (quantum cost) [7].

Various quantum image representation models have been proposed to utilize quantum mechanics to store and process images, most recently FRQI (Flexible Quantum Image Representation) [8] and NEQR (Novel Enhanced Quantum Image Representation). Although FRQI only uses a single qubit to store the greyscale color information for all pixels [9]. Therefore, some complex image operations cannot be done by FRQI. NEQR was proposed to improve the limitation of FRQI by employing two entangled qubit sequences to store the color and position information respectively [9]. Unlike with FRQI, we can store the whole image in the superposition of the two-qubit sequences and recover the accurate classical image from the NEQR quantum image model [9]. Later in 2016, Farouk et al. proposed an extension of NEQR called quantum multi-channel representation (QMCR) which can store multiple color channels into the entangled color sequence [10].

However, these quantum algorithms suffer from the complexity design of the required circuit for image representation. Therefore, we developed an algorithm based on [11] to reduce the overall number of Toffoli gates used to build the quantum circuit, but instead of using both the complement property and absorption rules of Boolean algebra we use only the former. We name our algorithm with TT-LITE (Ternary Tree LITE) since it uses only a single complement property of Boolean algebra. The performance analysis of our proposed algorithm does not degrade compared to [9] and provides a lower computational complexity versus tabular methods presented in [11]. Also, the required execution time and the space efficiency are improved compared to [12]. Therefore, the number of quantum gates is minimized and overall quantum cost is reduced. This will lead to design the quantum circuit in a feasible and efficient way for future physical implementation of quantum image processing applications.

The paper is structured as follows: Section **2** provides a brief introduction on quantum image preparation and discusses previous works in the field. Section **3** gives some background on the main idea behind quantum image compression, quantum gates and Toffoli gate ESOP representation. Subsection **3.1** introduces ternary trees as a data structure for ESOP representation, and **3.2** proposes our algorithm for compressing a quantum image circuit. Section **4** shows how we generate the quantum circuit after the ternary tree algorithm. Section **5** briefly introduces QISkit [4] and finally Sections **6** and **7** show experimental results and discuss the advantage of TT-LITE over previous methods.

## 2   Quantum Image Representation

Flexible representation for quantum images "FRQI" and novel enhanced quantum representation "NEQR" are the most important representations of quantum images. The architecture of FRQI consists of three main stages which are preparation the FRQI image state from the original one, compression, and processing operations as storing, detection and retrieving. The representation of quantum images using the flexible one is similar to the description of the pixel of images on classical computers. The process of representation is performed by capturing the vital information of both the colors and the similar positions for each pixel at each point in an image. Later, the resulted pixels integrated to form the quantum state as represented in **Eq. (1, 2)** as:

$$|I> = \frac{1}{2^n}\sum_{i=0}^{2^{2n}-1} |ci> \oplus |i> \qquad (1)$$

$$|ci> = \cos\theta_i |0> + \sin\theta_i |1> \qquad (2)$$

Where $|0>$ and $|1>$ denotes the two dimensional of quantum computational basis, $|i>, i = 1,2,3,\ldots 2^{2n-1}$, and $\theta$ represents a series of angles (vector) of the encrypted color. The flexible representation of quantum image consists of two main parts which $|c_i>$ and $|i>$. $|c_i>$ and $|i>$ encrypt the information regarding both the colors and its corresponding positions of the image. NEQR is developed to represent digital images by storing the grayscale value for each pixel by using the basis state of a qubit string rather than an angle encrypted in a qubit in FRQI. Accordingly, the whole image can be stored using two entangled qubit sequences, one for the grayscale and other for the positional information for each pixel.

The representative expression of a $2^n \times 2^n$ form of NEQR representation can be written as in **Eq. (3)**

$$|I> = \frac{1}{2^n} \sum_{y=0}^{2^n-1} \sum_{x=0}^{2^n-1} |f(y,x)> |yx> = \frac{1}{2^n} \sum_{y=0}^{2^n-1} \sum_{x=0}^{2^n-1} \oplus_{i=0}^{q-1} |c_{yx}^i> |yx> \qquad (3)$$

An illustrative example of a NEQR representation of a 2 × 2 quantum image and its corresponding state is shown in **Figure 1** and **Eq. (4, 5)**

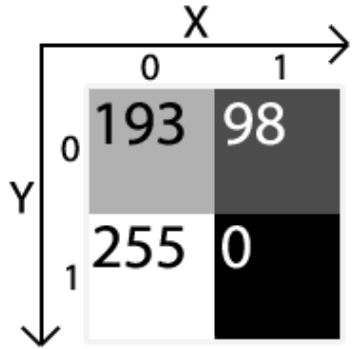

Figure 1 A 2x2 8-bit NEQR quantum image.

$$|I\rangle = \frac{1}{2}(|193\rangle \otimes |00\rangle + |98\rangle \otimes |01\rangle + |255\rangle \otimes |10\rangle + |0\rangle \otimes |11\rangle) \quad (4)$$

$$|I\rangle = \frac{1}{2}(|11000001\rangle \otimes |00\rangle + |11000010\rangle \otimes |01\rangle + |11111111\rangle \otimes |10\rangle + |00000000\rangle \otimes |11\rangle) \quad (5)$$

## 3  Quantum Image Compression

One of the main challenges of quantum image representation is reducing the number of quantum gates used to build the circuit of the image, also known as quantum image compression. Most of the quantum cost of an image circuit comes from the number of Toffoli gates needed to represent the image in a quantum system. A logic circuit consisting of Toffoli gates can be directly mapped to an exclusive sum of product "ESOP" function with a single output [14]. To generate the ESOP function, we look at each color line individually. For example, **Eq. 6** denotes the ESOP function associated with $C_0$ in **Figure 2** as:Error! Reference source not found.

$$f(C_0) = 00 \oplus 01 \oplus 10 \quad (6)$$

Toffoli gates consist of two parts of information: n-control bits and one target bit. In order to combine two Toffoli gates into one, the complement property rule of Boolean algebra: $ab + ab' = a–$ will be applied. Notice that $b$ and $b'$ have been removed and replaced with '–' which indicates "don't care". A don't-care value means we do not have any control at that target bit line and can be used if it will help in the process of minimization. Otherwise; it will be ignored. Minimization occurs when the complement rule is applied to these ESOP functions. **Eq. 7** is the result of minimization of **Eq. 6** after applying the complement property rule. **Figure 3** shows the minimized circuit.

$$f(C_0) = 0–  \oplus 10 \quad (7)$$

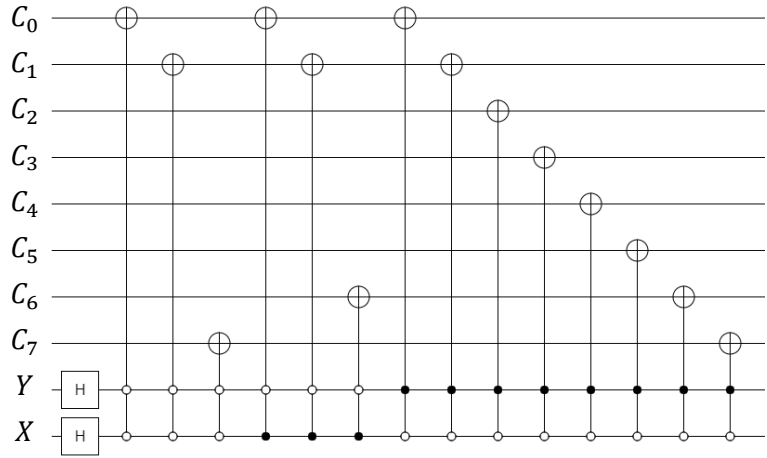

Figure 2 The quantum circuit of Figure 1

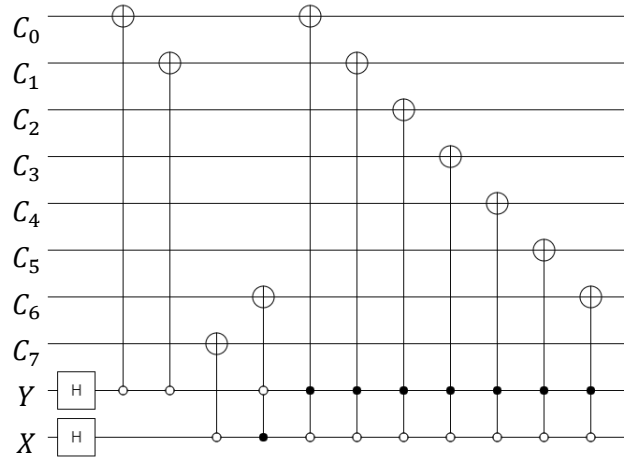

Figure 3 The result of minimizing all the color lines in Figure 2

### 3.1 Ternary Trees and Quantum Gates

The first approach of ternary trees to represent SOP (Sum of Product) is proposed in [15]. A ternary tree is like a binary tree, but where each non-terminal node may have up to three children (instead of 2 for the binary tree). Each non-terminal node has a parity of 0, – or 1 which translate to the values zero, "don't care" and one respectively. Therefore, the maximum size of a ternary tree is $O(3^n)$ since there are $3^n$ different terms for an $n$-input function [11], but the real ternary tree is usually much smaller. An illustrative example of the ternary tree representation for the ESOP function is shown in **Figure 4** and **Eq.8**.

$$f(x_1, x_2) = 0- \oplus 10 \oplus 11 \tag{8}$$

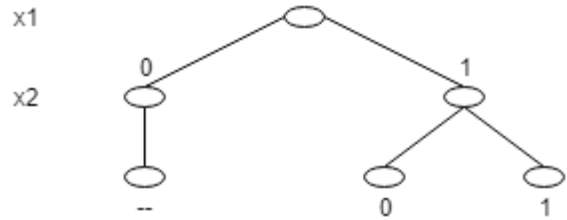

*Figure 4 Ternary tree ESOP representation of Eq. 8*

The representation of quantum images can be manipulated by applying the proper sequence of quantum gates. The quantum gate can work on a single or multiple qubits. In our proposed approach, minimization of the quantum circuit is achieved by using {*X, H, CNOT, Toffoli*} gates to minimize the required gates in the quantum image circuit. **Figure 5** shows a few variations of the NOT gate; **c)** is a Toffoli gate with two positive controls (represented as a filled circle), while **d)** has one positive and one negative control (represented as a hollow circle).

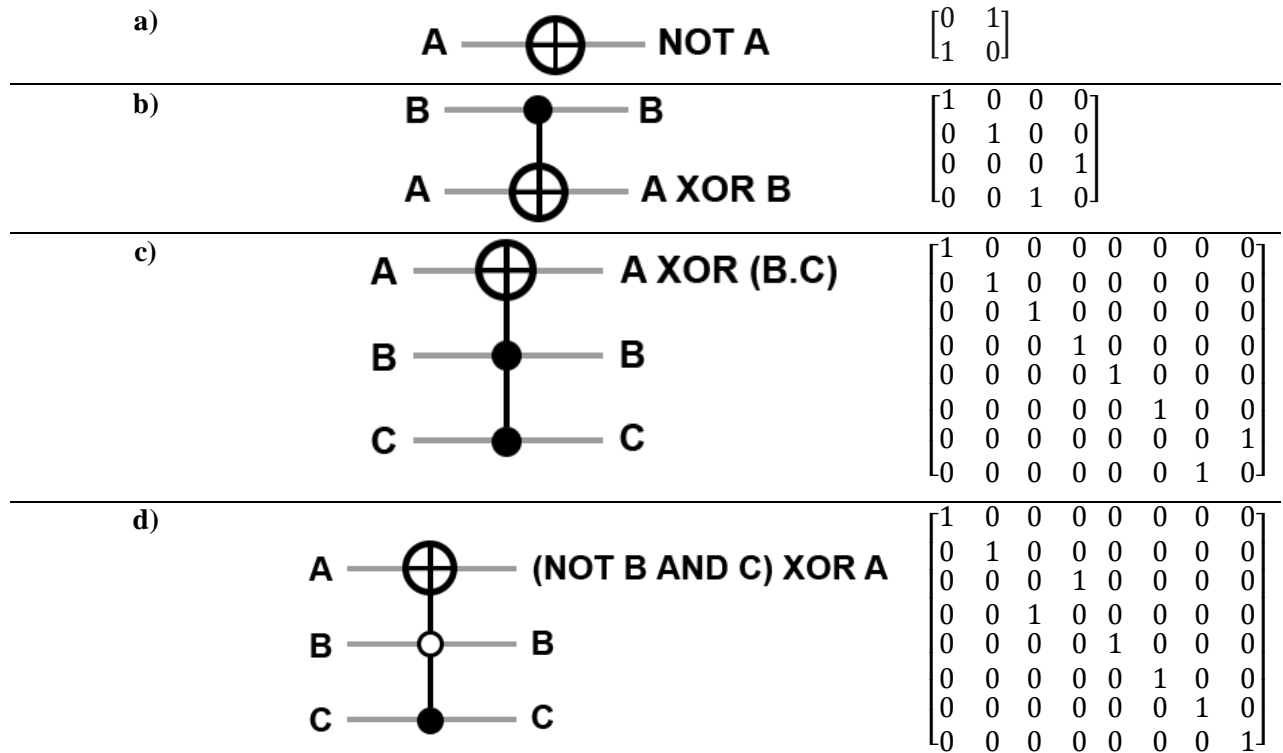

*Figure 5 a) NOT gate b) CNOT gate c) Toffoli gate with two positive controls d) Toffoli with one negative control and one positive control*

## 3.2 Ternary Tree ESOP Minimization Algorithm

Ternary tree ESOP minimization algorithm applies the complement rule of Boolean algebra only, targeting the reduction of the number of the ternary tree terminal nodes (leaves) [11]. Our protocol is a simplified version of the algorithm presented in [11] since we do not use the complement rule of Boolean algebra but using only some of the employed operations of [11, 15]. The algorithm consists of the following steps (see appendix A):

***STEP 1;*** if a non-terminal node at the $(n-1)$-th level has two successor nodes (which must be 0 and 1), then they always may be merged to obtain one "don't-care" terminal [11]. The resulting minimized tree after applying the absorption rule to **Figure 4** is shown in **Figure 6.**

Box 1: Merge Leaves of Node

```
MergeLeaves (node):
    if node is None or node is leaf:
        return

    if node has dc as None and lo and hi as leaf nodes:
        set lo and hi as None
        set dc as new node
    else:
        MergeLeaves(node.lo)
        MergeLeaves(node.dc)
        MergeLeaves(node.hi)
```

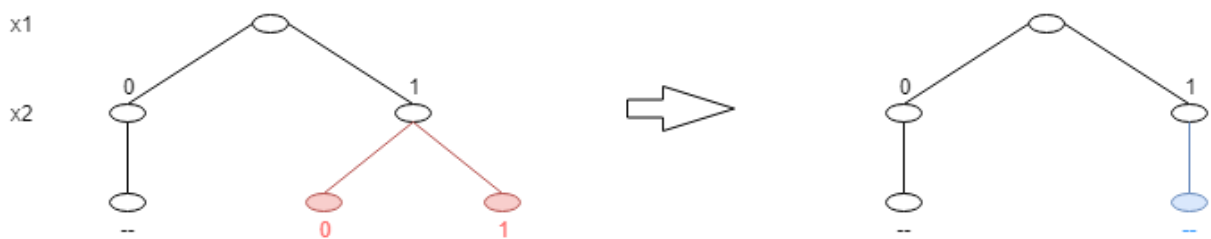

*Figure 6 Absorption rule on the two leaf nodes 0 and 1.*

It is easy to realize that the couple (10, 11) can be combined to 1–. The expression $f(x_1, x_2) = 0-\oplus 10 \oplus 11$ is now minimized to $f(x_1, x_2) = 0-\oplus 1-$. Since the reduction operations may be performed on the leaves of the tree only, no other tree reduction steps can be done at this stage; thus another phase of the minimization algorithm follows – the tree rotation [11].

***STEP 2;*** the tree rotation is achieved by cutting off the root node, which yields three separate trees (at most). Subsequently, the root variable is appended to all leaves of the trees [11]. **Figure 7** has only two trees after splitting because the root node of **Figure 6** has only two branches.

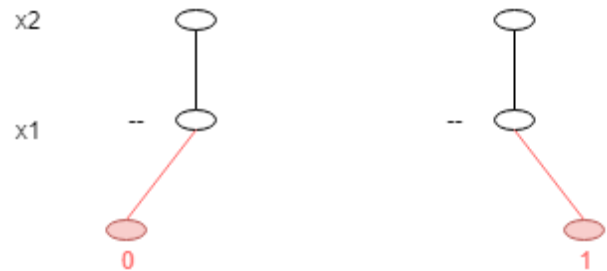

*Figure 7 The resulting tree after the split.*

Box 2: Merge Tress

```
MergeTrees (T1, T2):
      if not T1:
            return T2
      if not T2:
            return T1

      T1.lo = MergeTrees(T1.lo, T2.lo)
      T1.dc = MergeTrees(T1.dc, T2.dc)
      T1.hi = MergeTrees(T1.hi, T2.hi)

      return T1
```

Box 3: Append All Nodes

```
AppendAll (node, symbol):
      if not node:
            return

      if node is leaf:
            if symbol is 0:
                  node.lo = NewNode()
            if symbol is -:
                  node.dc = NewNode()
            if symbol is 1:
                  node.hi = NewNode()

      AppendAll(node.lo, symbol)
      AppendAll(node.dc, symbol)
      AppendAll(node.hi, symbol)
```

Box 4: Rotate Tree

```
Rotate (tree):
    if tree.lo:
        AppendAll(tree.lo, '0')
    if tree.dc:
        AppendAll(tree.dc, '-')
    if tree.hi:
        AppendAll(tree.hi, '1')

    return MergeTrees(tree.lo, MergeTrees(tree.dc,
    tree.hi))
```

*STEP 3;* the trees are merged together, by traversing these trees from their roots in parallel and merging nodes [11]. **Figure 8** is the result of merging the two trees of **Figure 7**.

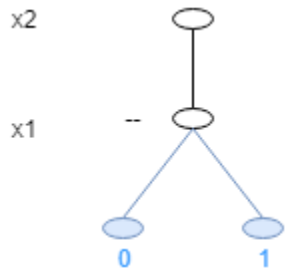

*Figure 8 The merged tree after the split.*

Since, the bottom leaf nodes can be merged; *The algorithm* consists of the following steps (see appendix A):

*STEP 1* is repeated to obtain a function with no minterm and only "don't-care". Usually, the result is a combination of minterms and "don't-care". To make sure that every minterm is looked at, steps 1-3 are repeated such that the tree is rotated n-times (where n is the number of variables). We explore each color line $C_i$ where $i = [0, q]$ of the quantum image circuit and use the control masks for all Toffoli gates on color line $C_i$ as the input for our algorithm. The input for color line $C_0$ in **Figure 2** and for the TT-LITE algorithm is [00, 01, and 10]. The appropriate ternary tree data structure is then constructed.

Box 5: Minimize Function

```
Minimize (F):
    tree = CreateTernaryTree(F)

    for range(i):
        MergeLeaves(tree)
        tree = Rotate(tree)
```

Minimize

return Traverse(tree)

## 4 Toffoli Generation

For most of the current quantum image circuits, some of the Toffoli gates in the circuit will have negative polarity controls. In this case, the control bit has to be flipped before it interacts with a toffoli containing a negative polarity control. This is accomplished by retrieving the output from the ESOP minimization algorithm of the current color line and input the results into our toffoli generation algorithm. We use an array $X[i]$ initialized to 0, where $0 < i < h + w$, to keep track of the current X-gates where $X[i] = 0$ if no X-gate is acting on location line $L_i$, otherwise, $X[i] = 1$. We start by getting the output of the TT-LITE algorithm for the current color line $C_j$ where $0 < j < q$. Suppose we have a toffoli gate $TOF_{L_i}^n$, where $0 < n < \#\, toffolis\ on\ color\ line\ C_j$ and $TOF$ have polarity $[0, -, 1]$ acting on location line $L_i$. We iteratively check each control bit in $TOF_{L_i}^n$. There are only two cases:

$if\ X[i] = 0\ and\ TOF_{L_i}^n = 0$:

$\quad add\ X - gate\ on\ L_i\ and\ mark\ X[i] = 1$ or

$if\ X[i] = 1\ and\ TOF_{L_i}^n = 1$:

$\quad add\ X - gate\ on\ L_i\ and\ mark\ X[i] = 0$

The second if statement reverses the action of the first. Suppose that a quantum image circuit where $L_i$ is a control line, and $C_j$ is a target line is represented in **Figure 9.**

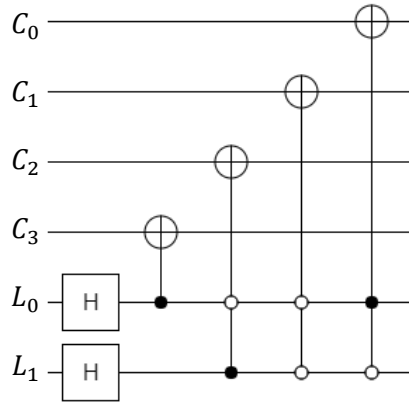

Figure 9 A quantum image circuit.

Examination the first Toffoli (left to right), we have $TOF_{L_0}^0 = 1\ and\ X[0] = 0$ so this doesn't meet either of our conditions. The second Toffoli $[TOF_{L_0}^1 = 0, TOF_{L_1}^1 = 1]$ has a negative polarity control on $L_0$. So, we add an X-gate to $L_0$ and mark $X[0] = 1$. The third Toffoli $[TOF_{L_0}^2 = 0, TOF_{L_1}^2 = 0]$ has two negative polarity controls. Since $TOF_{L_0}^2 = 0\ and\ X[0] = 1$, in this case we can skip the first control because there is already an X-gate acting on $L_0$. The second control meets our condition $TOF_{L_1}^2 = 0 \text{ and } X[1] = 0$, and so we add an X-gate on $L_1$ and mark $X[1] = 1$. The last Toffoli $[TOF_{L_0}^3 = 1, TOF_{L_1}^3 = 0]$ has one positive and one negative control. The first control $TOF_{L_0}^3 = 1 \text{ and } X[0] = 1$ meets our second condition, and so we must reverse the action of the previous X-gate by adding another X-gate to $L_0$ and mark $X[0] = 0$. Once all of the Toffoli gates have been generated for each color line, we add an X-gate on any remaining $X[i] = 1$. The resulting quantum image circuit after applying the X-gate

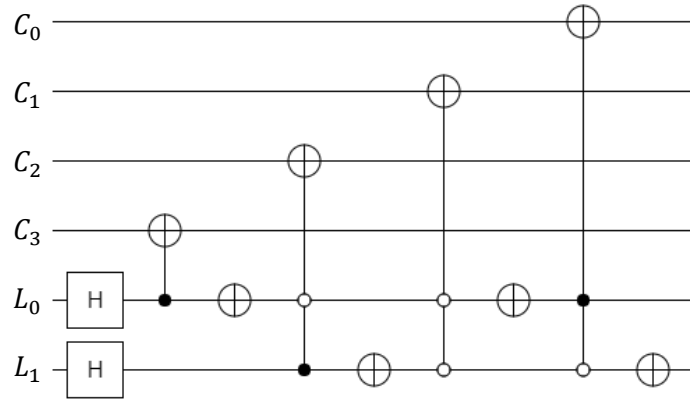

Figure 10 The resulting quantum image circuit after applying the X-gate algorithm.

algorithm is shown in **Figure 10**.

Although this method is good enough for experimental purposes, we also require a better approach of efficiently cascading Toffoli gates to reduce the overall quantum cost of the X-gates. Therefore, implementing Toffoli gate cascade generation [16] would achieve better ordering of the Toffoli gates and thus reduce the number of X-gates needed alongside mixed polarity Toffoli gates.

## 5   QISkit Implementation

QISkit (Quantum Information Science kit) is IBM's open-source quantum API implemented in Python [4]. It allows us to interact with a real quantum computer and perform experiments. A circuits "depth" is the number of physical gates used. The purpose of using QISkit is to record the depth of the circuit after it has been made suitable to run on a real quantum computer. One problem we came across when building the quantum image circuit was how to represent an n-bit Toffoli gate in QISkit. To implement an $n$-bit Toffoli gate, where $n > 2$, we need $n - 1$ ancilla qubits, $2(n - 1)$ toffoli gates and 1 control $U$ operation (the control $U$ operation is a CNOT for our case because we want to achieve an $n$-bit Toffoli), where $U$ is an arbitrary single-control gate. This is not taken into consideration for the Toffoli count in the experimental results section; we only use this to realize an $n$-bit Toffoli gate using QISkit compiled in Python 3.5. **Figure 11** shows a general control $U$ operation for the case $n = 5$ [17]. The depth results are recorded in **Table 1** under *Overall Gate Count*.

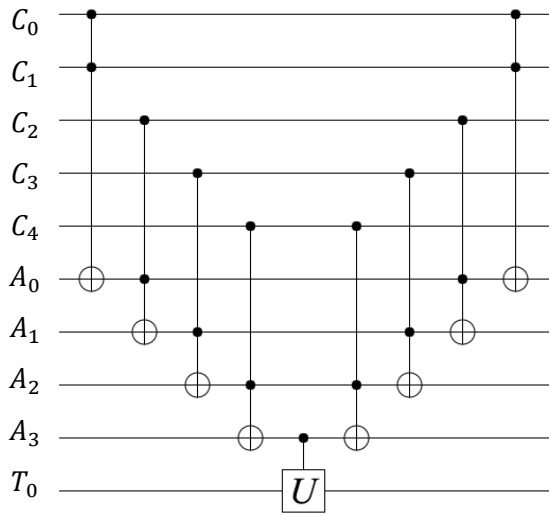

Figure 11 The implementation of the $C^n(U)$ operation for the case $n = 5$

## 6   Experimental Results

All experiments were conducted on an Intel Core i7-7700K @ 4.4GHz, 16GB DDR4 RAM on Windows 10 using Python 3.5. The result of minimization on 7 greyscale images of **Figure 12** is shown in **Table 1** . It contains the initial Toffoli gate count is along with the reduced Toffoli count after running TT-LITE on the image. The compression ratio shows the relationship between the two counts and CPU time for TT-LITE is also shown with an overall gate count. The result comes from implementing the resulting TT-LITE image circuit in QISkit [4] and counting the overall number of gates required by IBM's physical quantum computer. The compression rate is ranging from 64.25 to 99.98%, Horizontal.png is especially interesting, achieving a near perfect compression ratio of 99.98% .

 **Table** 2 and **Figure 13** compare TT-LITE results against [9] where the sum-of-product minimization algorithm Espresso [12] is used. we observe that the average compression result is almost identical to that of Espresso. The main weakness with Espresso is downgrading of performance when the number of terms increased and reached thousands of terms [12]. This is not ideal for image processing as images are realistically large with multiple color channels. The main advantage of TT-LITE is that its performance does not degrade like Espresso [12] and clearly achieves similar results.

**Figure 14** represents an RGB image scaled from 200x200 pixels down to 10x10 pixels in 10-pixel increments and **Figure 15** shows the performance of TT-LITE on these RGB images. For larger quantum images, the number of quantum gates increases exponentially. With ~450,000 gates TT-LITE reduces it to ~160,000 Toffoli gates in 30 seconds. This demonstrates the speedup TT-LITE has when compared to Espresso [12].

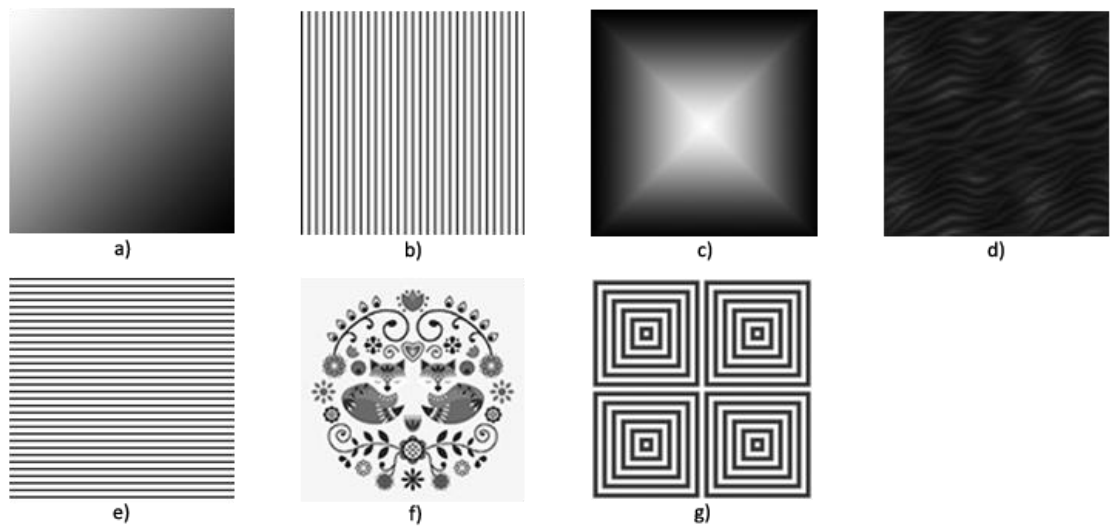

*Figure 12  Various 128px x 128px images. a) gradient.png  b) vertical.png  c) star.png  d) pattern1.png  e) horizontal.png  f) manmade.png  g) pattern2.png*

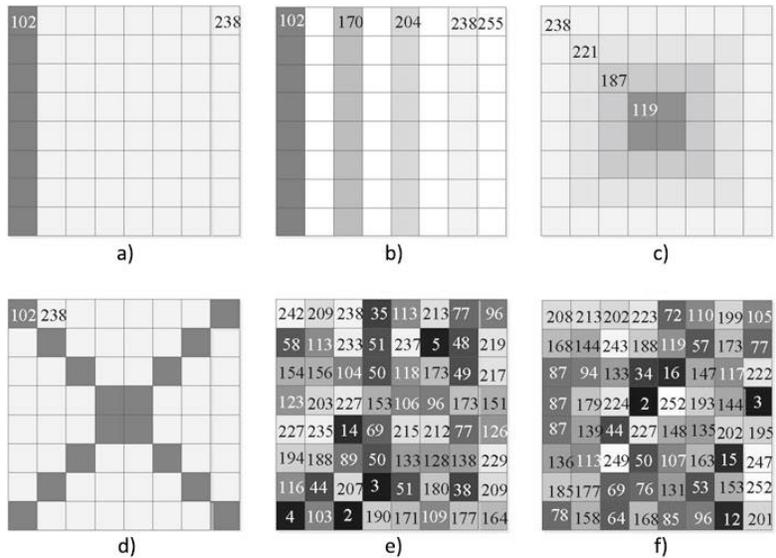

*Figure 13 Test case images taken from [9]. a) to d) are somewhat regular while e) and f) are random.*

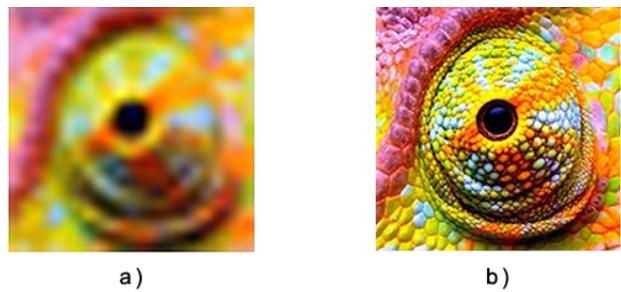

*Figure 14 An RGB image a) 10px x 10px (scaled up) b) 200px x 200px.*

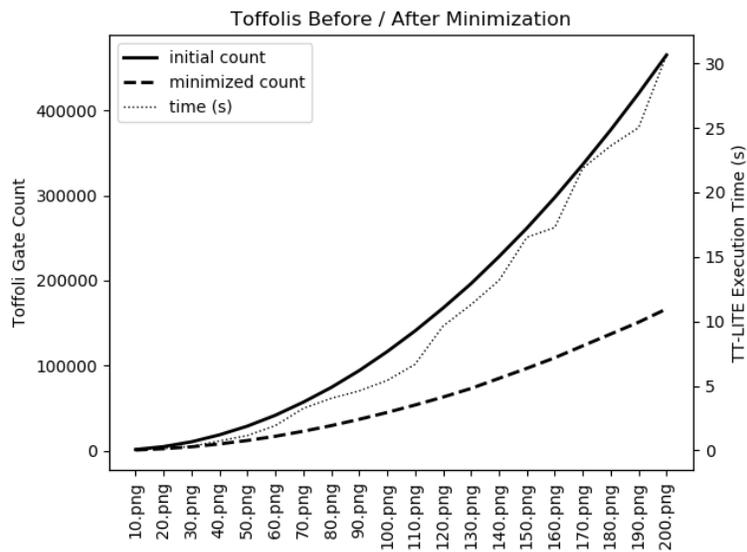

*Figure 15 Test results on the RGB image in Figure from 10px x 10px scaled all the way to 200px x 200px.*

*Table 1 Test results of TT-LITE on images in Error! Reference source not found..*

| Image File | Initial Toffoli Count | Minimized Toffoli Count | Compression Ratio (%) | Time (s) | Overall Gate Count |
|---|---|---|---|---|---|
| gradient.png | 65810 | 17872 | 72.84 | 2.5601 | 490635 |
| vertical.png | 97408 | 244 | 99.75 | 1.3404 | 3283 |
| horizontal.png | 98304 | 16 | 99.98 | 1.2088 | 69 |
| star.png | 57812 | 11127 | 80.75 | 1.8540 | 297510 |
| pattern1.png | 49020 | 17527 | 64.25 | 2.3058 | 485244 |
| manmade.png | 90371 | 19844 | 78.04 | 2.6449 | 536399 |
| pattern2.png | 79416 | 3460 | 95.64 | 1.5608 | 78895 |

*Table 2 Test results of TT-LITE compared to results given in [9] on images in Error! Reference source not found..*

|  | a) | b) | c) | d) | e) | f) | AVG (%) |
|---|---|---|---|---|---|---|---|
| *FRQI* | 93.75 % | 92.19 % | 50 % | 62.5 % | 0% | 1.56 % | 50 |
| *Gate Count* | 4/64 | 5/64 | 32/64 | 24/64 | 64/64 | 63/64 |  |
| *NEQR (Espresso)* | 97.28 % | 95 % | 78.13 % | 85.23 % | 45.35% | 43.65 % | 74.11 |
| *Gate Count* | 10/368 | 20/400 | 84/384 | 52/352 | 141/258 | 142/252 |  |
| *TT-LITE* | 98.37 % | 95 % | 68.75 % | 86.36 % | 50.96 % | 50 % | 74.90 |
| *Gate Count* | 6/368 | 20/400 | 120/384 | 48/352 | 127/259 | 127/254 |  |

## 7  Conclusion

TT-LITE – an algorithm for minimizing Toffoli gates in quantum image circuits is proposed. Ternary trees are used to minimize exclusive sum of product terms. As a result, the runtime of the image compression step in quantum image preparation is reduced when compared to Espresso that is used in [9]. It was experimentally shown that when given thousands of product terms Espresso takes longer CPU time than this method. Another advantage is that this method is specifically tailored to the quantum image compression problem using only a single rule when minimizing Toffoli gates, while Espresso is tailored to any SOP minimization problem which attributes to its runtime complexity. It was shown that this method achieves similar results to that of Espresso in seconds, making it ideal for image processing.

**Acknowledgement**

This research has been supported by the Natural Sciences and Engineering Research Council of Canada (NSERC) and contribution of Undergraduate Student Research Award (USRA) under the supervision of Dr. Shohini Ghose. Special thanks to Ahmed Farouk for guidance and assisting in editing this paper.